\documentclass[twocolumn, twocolappendix]{aastex631}

\usepackage{soul}
\usepackage{upgreek} 
\usepackage{hyperref}

\received{May 19, 2023}
\accepted{September 05, 2023}

\submitjournal{ApJL}

\shorttitle{Correlating Neutrinos with 5BZCAT and RFC}
\shortauthors{Bellenghi et al.}
\graphicspath{{./}{}}

\begin{document}

\title{Correlating high-energy IceCube neutrinos with 5BZCAT blazars and RFC sources}

\correspondingauthor{Chiara Bellenghi}
\email{chiara.bellenghi@tum.de}

\author[0000-0001-8525-7515]{Chiara Bellenghi}
\affil{Technical University of Munich\\ TUM School of Natural Sciences, Department of Physics\\
James-Franck-Straße 1,
D-85748 Garching bei M\"unchen, Germany}

\author[0000-0002-4707-6841]{Paolo Padovani}
\affil{European Southern Observatory \\ Karl-Schwarzschild-Straße 2 \\
D-85748 Garching bei M\"unchen, Germany}

\author[0000-0003-0705-2770]{Elisa Resconi}
\affil{Technical University of Munich\\ TUM School of Natural Sciences, Department of Physics\\
James-Franck-Straße 1,
D-85748 Garching bei M\"unchen, Germany}

\author[0000-0002-2265-5003]{Paolo Giommi}
\affil{Institute for Advanced Study -
Technische Universit{\"a}t M{\"u}nchen\\
Lichtenbergstrasse 2a, D-85748 Garching bei M\"unchen, Germany}
\affil{Center for Astro, Particle and Planetary Physics (CAP3)\\ 
New York University Abu Dhabi\\ 
PO Box 129188, Abu Dhabi, United Arab Emirates}
\affil{Associated to INAF, 
Osservatorio Astronomico di Brera\\
via Brera, 28, I-20121 Milano, Italy}



\begin{abstract} 
We investigate the possibility that blazars in the Roma-BZCAT Multifrequency Catalogue of Blazars (5BZCAT) are sources of the high-energy astrophysical neutrinos detected by the IceCube Neutrino Observatory, as recently suggested by \cite{Buson2022a,Buson2022b}. Although we can reproduce their $\sim 4.6\, \sigma$ result, which applies to 7 years of neutrino data in the Southern sky, we find no significant correlation with 5BZCAT sources when extending the search to the Northern sky, where IceCube is most sensitive to astrophysical signals. To further test this scenario, we use a larger sample consisting of 10 years of neutrino data recently released by the IceCube collaboration, this time finding no significant correlation in either the Southern or the Northern sky. These results suggest that the strong correlation reported by \cite{Buson2022a,Buson2022b} using 5BZCAT could be due to a statistical fluctuation and possibly the spatial and flux non-uniformities in the blazar sample. 
We perform some additional correlation tests using the more uniform, flux-limited, and blazar-dominated Radio Fundamental Catalogue (RFC) and find a $\sim 3.2\sigma$ equivalent p-value when correlating it with the 7-year Southern neutrino sky. However, this correlation disappears completely when extending the analysis to the Northern sky and when analyzing 10 years of all-sky neutrino data. Our findings support a scenario where the contribution of the whole blazar class to the IceCube signal is relevant but not dominant, in agreement with most previous studies. 
\end{abstract}



\section{Introduction} \label{sec:intro}
The quest for the sources of the highest energy cosmic radiation is closely connected with the search for the production sites of high-energy astrophysical neutrinos, generated in cosmic ray interactions.
In this pursuit, the IceCube Neutrino Observatory\footnote{https://icecube.wisc.edu/}\citep{IceCubeJInst2017} has been playing a leading role for over a decade, from the detection of hundreds of astrophysical neutrinos with energies reaching above the PeV ($10^{15}$~eV), which marked the birth of high-energy (TeV -- PeV) neutrino astrophysics \citep{IceCube2013}, to the recently reported evidence for neutrino emission from the Seyfert II galaxy NGC~1068 \citep{IceCubeScience2022}.
The first association with an extragalactic source of high-energy neutrinos happened in September 2017, when IceCube detected an event with very high energy ($\sim$ 290~TeV) and a high probability of having astrophysical origin.
Follow-up multi-messenger observations by the {\it Fermi}-LAT\footnote{https://fermi.gsfc.nasa.gov/science/instruments/lat.html} and MAGIC\footnote{https://magic.mpp.mpg.de/} collaborations confirmed that the neutrino event had a $\gamma$-ray counterpart in the blazar\footnote{Blazars are active galactic nuclei (AGN) whose emission is dominated by a relativistic jet pointing towards the observer: e.g. \cite{Urry1995,Padovani2017}.} TXS~0506+056, which was also exhibiting an enhanced $\gamma$-ray activity at the time of the neutrino detection \citep{txs2018}.
The chance coincidence probability of correlated neutrino and $\gamma$-ray emissions was found to be at the 3$\sigma$ level. An a-posteriori analysis of IceCube archival data at the position of TXS~0506+056 detected a neutrino flare between September 2014 and March 2015 incompatible with the background at the 3.5$\sigma$ level \citep{IceCube2018}.
The TXS~0506+056 association immediately triggered great interest in the hypothesis of a correlation between blazars and astrophysical neutrinos, which had been already suggested in several studies antecedent to the September 2017 event \citep[e.g.][and references therein]{PadovaniResconi2014,Padovani2016,Lucarelli2017,IceCubeICRC2017}.
Since then, many studies have addressed this topic trying to find statistically robust associations, mostly with inconclusive ($\lesssim 3\sigma$) results \citep[e.g., see the review by][and references therein]{Giommmi2021,Abbasi2023}.
\cite{Buson2022a} have recently cross-matched the 5th edition of the Roma-BZCAT Multifrequency Catalogue of Blazars (5BZCAT: \citealt{BZCAT5}) with a neutrino p-value map covering 7 years of observations made by IceCube between 2008 and 2015 \citep{7yearPS}.
The neutrino sky published by IceCube maps the local probability of the observation to be consistent with the background only, at each location in the sky.
Due to its unique location at the geographic South Pole,
neutrinos coming from the Northern sky have to cross the Earth to reach the IceCube detector. Above PeV energies, they start to be absorbed in the Earth at declinations $\delta>30^{\circ}$. Therefore, \cite{Buson2022a} focused only on the Southern sky (i.e., $-85^{\circ} < \delta < -5^{\circ}$), arguing that this region is the most sensitive to an astrophysical signal in the PeV---EeV energy range. They came to the conclusion that blazars were associated with high-energy astrophysical neutrinos with a post-trial chance probability of $6 \times 10^{-7}$, later scaled down to $2 \times 10^{-6}$ \citep{Buson2022b}, equivalent to $\sim 4.6\sigma$.
The main purpose of this paper is to test the results of \cite{Buson2022a,Buson2022b}, expand their study first to the 7-year Northern sky, and then to the whole sky using a recently released, $\sim 60\%$ larger and improved, IceCube dataset covering 10 years of observations \citep{ICdatarelease}. 

\section{The blazar samples}\label{blazar_sample}
5BZCAT contains coordinates and multi-frequency data for 3561 sources, either confirmed blazars or exhibiting blazar-like characteristics.
When it was first released, this {\it list} of blazars was likely the most comprehensive one available, as it included the vast majority of sources well documented in the literature.
By construction, however, 5BZCAT is a very inhomogeneous sample, as it collects sources detected in various frequency bands and different surveys, and is not flux-limited, i.e., it does not include all objects above a certain flux in a given band.
Following \cite{Buson2022a}, we exclude sources overlapping with the Galactic plane (galactic latitude $|\rm b_{II}| < 10^{\circ}$), consider only $|\delta| < 85^{\circ}$, and do not include the 92 objects classified as ``BL Lac candidates'' in 5BZCAT.

To assess the impact of using a flux-limited and uniformly distributed sample, as compared to an inhomogeneous list of targets, we also use a subset of the very-long baseline radio interferometry (VLBI)-based Radio Fundamental Catalogue\footnote{http://astrogeo.org/rfc/} (RFC), that is a radio-selected sample of 3411 AGN with an average 8 GHz VLBI flux density $\ge 150$ mJy used by \cite{Plavin2021}. This was originally selected by \cite{Plavin_2020}, who noted that a special effort was made to ensure completeness above this limit. (As shown in \autoref{sec:discussion_RFC}  
the RFC sources are distributed more uniformly in the sky than the ones in the 5BZCAT, especially in the Northern sky).
For consistency, we use here the same version as in the latter paper (rfc\_2020b), which tested for an association between the RFC and IceCube neutrinos (see \autoref{sec:discussion_RFC}). 
As done for 5BZCAT, we exclude RFC sources with $|\rm b_{II}| < 10^{\circ}$ and $|\delta| > 85^{\circ}$. The RFC is dominated by Doppler-boosted AGN, i.e., blazars. In fact, when 
cross-correlating it with a ``master blazar sample'' composed of the blazars in the 5BZCAT, 3HSP \citep{Chang_2019}, 
and {\it Fermi} 4FGL-DR3 \citep{4LAC-DR3} catalogs, which includes 6,425 unique sources, we 
find ~59\% matches. Since not all blazars in the Universe have been identified, 
this provides a robust lower limit to the RFC blazar fraction. This percentage  
increases to $\gtrsim 84$\% when we make a cut at 0.5~Jy instead of 0.15~Jy, 
which shows that brighter radio blazars are more likely to have been identified.


\section{The 10-year neutrino sample}\label{nu_sample}
The IceCube collaboration has recently released 10 years of candidate neutrino events detected between April 6, 2008 and July 10, 2018 \citep{ICdatarelease}\footnote{https://icecube.wisc.edu/data-releases/2021/01/all-sky-point-source-icecube-data-years-2008-2018/}.
The sample is optimized for the search of neutrino point sources, thus including track-like events,
primarily due to muon-neutrino candidates (muon tracks), from all directions in the sky.
In comparison to the previous 7-year sample \citep{7yearPS}, utilized to generate the neutrino p-value sky map used by \cite{Buson2022a}, the number of events is increased by $\sim60\%$, and an updated processing procedure and a new angular reconstruction are applied to the events recorded between April 2012 and May 2015 (and to the events recorded afterward). Notably, the enhanced directional reconstruction improved the average angular resolution by more than 10\% for events with reconstructed muon energies above 10~TeV \citep{10yearPS,2019PhDCarver}. In the Northern sky, the range of reconstructed energies goes from a few tens of GeV to $\sim$1~PeV, with most events reconstructed at 1~TeV. The Southern sky, instead, covers a wider spectrum of energies, going up to 10~PeV.
Indeed, the absorption of the highest-energy neutrinos in the Earth reduces the detection efficiency of IceCube for neutrino energies above 1~PeV at $\delta\gtrsim30^{\circ}$.
On the other hand, the Southern neutrino sky is dominated by the background of muons and muon bundles created in atmospheric cosmic ray showers --- that resemble high-energy tracks \citep{IceCubeAP2016} ---, which are completely absorbed above $\delta=5^{\circ}$.
These considerations leave the horizon ($-5^{\circ}\lesssim\delta\lesssim30^{\circ}$) as the region where IceCube is the most sensitive to an astrophysical flux of neutrinos.

The 1,134,450 events in the released 10-year sample were previously used by the IceCube collaboration for a time-integrated search for neutrino point sources in both hemispheres \citep{10yearPS}.
A machine-readable version of the neutrino p-value map in \cite{10yearPS} is not publicly available. However, the data release includes the experimental data with their per-event measured physical properties (time, direction, and energy), the binned effective areas, and the binned instrument responses for 5 data-taking seasons, both not provided per-event. The effective areas are published as two-dimensional histograms in neutrino energy and declination, while the detector response functions are provided as five-dimensional histograms mapping a neutrino to the probability of the reconstructed observables in the detector. With this information {\bf it} is possible to analyze the data to search for astrophysical neutrino emission.
We produce our own neutrino p-value map with the public 10-year sample using an unbinned maximum likelihood-ratio method, following the prescription outlined in \cite{Braun2008} (see \autoref{app:llh} for details).
However, it is worth mentioning that only three bins for the true incoming neutrino declination are provided in the detector response matrices, covering the sky from $-90^{\circ}$ to $-10^{\circ}$, $-10^{\circ}$ to $+10^{\circ}$, and $+10^{\circ}$ to $+90^{\circ}$, respectively.
The coarseness of this information causes a decrease of $\sim$20--50\% in the sensitivity of the analysis to an astrophysical neutrino flux with respect to what is published in \cite{10yearPS}. This discrepancy may be attributed to the provided detector response being averaged over wide declination ranges, where it undergoes considerable variations (see \autoref{app:skymap}).
It should also be noted that the likelihood formalism adopted in this work does not exactly conform to the one used by the IceCube collaboration (see \autoref{app:llh} for details on the differences).
Therefore, the neutrino p-value map based on the 10-year data release produced within this work should not be considered an exact reproduction of the one published in \cite{10yearPS}, but should instead be seen as the outcome of a likelihood-based search for neutrino sources which finds results mostly statistically compatible to what has been published by the IceCube Collaboration (see \autoref{app:skymap}).

From the 10-year neutrino p-value map that we obtain from the likelihood scan of the sky, we can then select the most interesting neutrino spots for the cross-correlation analyses described in the following section.

\section{Statistical analysis method}\label{sec:method}
We perform various positional correlation analyses similar to the one described in \cite{Buson2022a}.
Their test correlates the sources in 5BZCAT (see \autoref{blazar_sample}) with the Southern neutrino sky {\bf by using} 7 years of muon tracks recorded by the IceCube detector \citep{7yearPS}. 
These have been published in the form of pre-trial local p-values on a grid of equivalent pixels of $\sim (0.11^{\circ})^2$ constructed using a HEALPix \citep{Gorski2005} projection with \textsc{Nside} parameter 512\footnote{https://icecube.wisc.edu/data-releases/2020/02/all-sky-point-source-icecube-data-years-2012-2015/}.
We follow \cite{Plavin2021} and \cite{Buson2022a} and denote the negative logarithm of local p-values mapping the neutrino sky as $L = -\log_{10} p$ to avoid confusion with the p-values of the correlation analyses.
The $L$ values are based on a maximum likelihood-ratio comparing a background and a signal hypothesis, where in the former case the neutrino emission is due to atmospheric 
background and diffuse astrophysical emission, while in the latter an additional component comes from the source and clusters around it.
Hence, a larger value of $L$ indicates a lower probability that the neutrino data are consistent with the background-only hypothesis (see \autoref{app:llh}).

We extend the analysis to the 7-year Northern sky.
Furthermore, we search for correlations between the same neutrino sample and the RFC sources (see \autoref{blazar_sample}).
Finally, we increase the size of the neutrino sample by using an updated neutrino skymap based on the 10-year data release (see \autoref{nu_sample}) and search for correlations between the newer neutrino sample and the two aforementioned catalogs.

As we want to follow the same statistical method developed in \cite{Buson2022a}, we also search for a spatial correlation between the blazars and the IceCube neutrinos using the hotspots in the neutrino sky map of $L$ values.
The hotspots are defined as all independent pixels with local significance $L$ above some pre-defined threshold $L_{\mathrm{min}}$. For this type of search, the hotspots have to be selected so as not 
{\bf to} double-count possible interesting neutrino signals.
In fact, the pixels in a skymap are necessarily correlated because of the same events contributing to the likelihood at neighboring locations.
To identify independent hotspots, we select pixels above the chosen significance threshold whose centers are at least $1^{\circ}$ apart from each other \citep{7yearPS,8yearPS}.
The test statistic (TS) of the correlation analysis is the number of hotspots with significance above a certain local threshold $L_{\rm{min}}$ having at least one source closer than some association radius $r_{\mathrm{assoc}}$.
A set of $L_{\rm{min}}$ and $r_{\mathrm{assoc}}$ values is defined a priori and a TS is computed for each possible combination.
For each pair of the $L_{\rm{min}}$ and $r_{\mathrm{assoc}}$ parameters, the p-value of the correlations is defined as the chance probability to get the observed TS value. We derive it by running the correlation analysis on pseudo-experiments generated by randomizing the position of the sources within $10^{\circ}$ from their original position. In this way, the original structure of the catalog is preserved in the simulated ones (see \citealt{Buson2022a} for more details).
The combination of the analysis parameters yielding the smallest p-value is the best-fit result.
The best p-value then needs to be corrected for the ``look-elsewhere effect'', i.e. for having tested all possible combinations of $L_{\rm{min}}$ and $r_{\mathrm{assoc}}$.
We choose different values of the analysis parameters depending on the neutrino sample that is being tested, as detailed in the following.

\subsection{Correlation with the 10-year neutrino sample}\label{sec:corr_10yr}

\begin{figure}[t!]
    \centering
    \includegraphics[width=\linewidth]{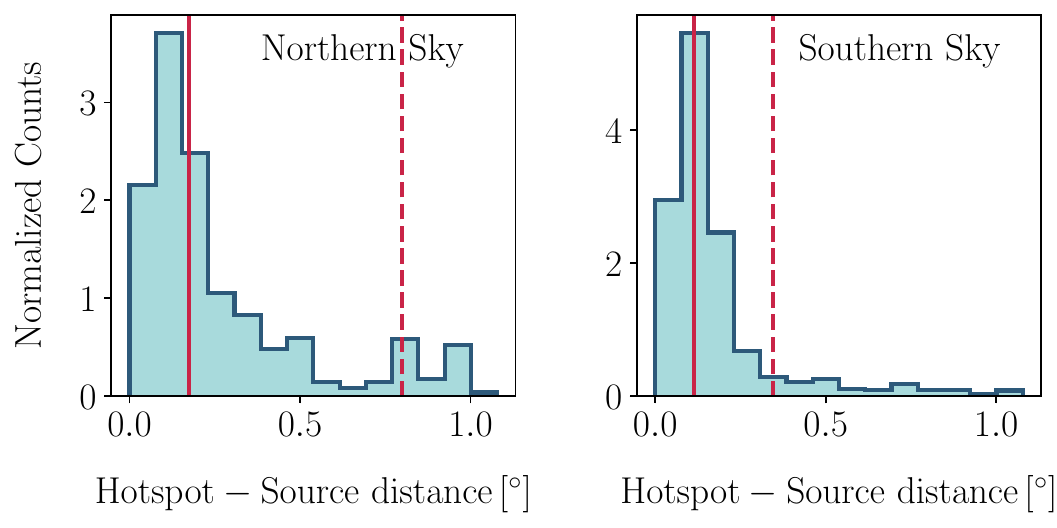}
    \caption{Distributions of the angular separation between the location of a simulated neutrino source and the associated hotspot with $\bf L > 3.0$ for the Northern (left) and Southern (right) sky. In both panels, the solid and dashed red lines indicate the median and the 90\% quantile of the distribution, respectively.}
    \label{fig:localization}
\end{figure}
For the correlation test using the 10-year neutrino $L$ map, we redefine the analysis parameters $L_{\rm{min}}$ and $r_{\mathrm{assoc}}$.
Given the reduced sensitivity of our point source analysis compared to the one published in \cite{10yearPS}, we decided to start scanning the $L_{\rm{min}}$ parameter from a lower significance of 3.0, instead of 3.5, in bins of 0.5 up to 4.5.
As for $r_{\mathrm{assoc}}$, we chose its range based on the capability of our point source likelihood analysis to localize a neutrino source in the sky.
To estimate this localization performance, we simulate astrophysical neutrino sources emitting with an E$^{-2}$ energy spectrum at various locations in the sky. For each simulation, we search the associated hotspot in a circle of $1^{\circ}$ radius centered at the source location and calculate the angular distance between the two.
\autoref{fig:localization} displays the distribution of these angular distances.
We do it separately in the Northern and Southern sky as we expect different outcomes due to the different energy ranges and data selection procedures used for the two hemispheres \citep{ICdatarelease}.
To simulate the sources, signal events are injected by sampling them from the detector response matrices, which map the true incoming neutrino energy and declination to the distribution of reconstructed muons that the detector would see.
This procedure intrinsically takes into account the smearing due to the detector resolutions for direction and energy reconstructions.
To locate the neutrino hotspot, we divide the circle around the source into pixels of $\sim (0.11^{\circ})^2$, maximize the likelihood-ratio (see \autoref{app:skymap}), and calculate the local significance $L$ at each of them.
The pixel with the maximum local $L$ value is taken as the hotspot for that simulation.
By repeating this procedure $\sim$ 10,000 times in each hemisphere, we find that 50\% (90\%) of the hotspots with $L>3.0$ are localized closer than $0.17^{\circ}$ ($0.80^{\circ}$) from the injection location in the Northern sky\footnote{The long, high angular separation tail of the distribution in the Northern sky is caused by the poor directional reconstruction quality of events with very low reconstructed muon energy. Indeed, the average separation between the parent neutrino and the reconstructed muon is $\sim 0.9^{\circ}$ \citep{2019PhDCarver} and the uncertainty on the reconstructed direction peaks at $\gtrsim1^{\circ}$ for events with neutrino energy below the TeV \citep{ICdatarelease}.}, while in the Southern sky 50\% (90\%) of the selected hotspots are closer than $0.13^{\circ}$ ($0.35^{\circ}$) from the simulated source position (see \autoref{fig:localization}).
A smaller association region corresponds to a reduced chance probability of associating a neutrino hotspot with a source.
Hence, to always include at least 50\% of the distributions and maximally reduce the background, we scan $r_{\mathrm{assoc}}$ from $0.20^{\circ}$ and go up to $0.70^{\circ}$ (the maximum $r_{\rm{assoc}}$ values used by \citealt{Buson2022a}) in steps of $0.05^{\circ}$.

\section{Results}

\subsection{5BZCAT -- 7-year hotspots correlation}

\begin{figure}[t!]
    \centering
    \includegraphics[width=\linewidth]{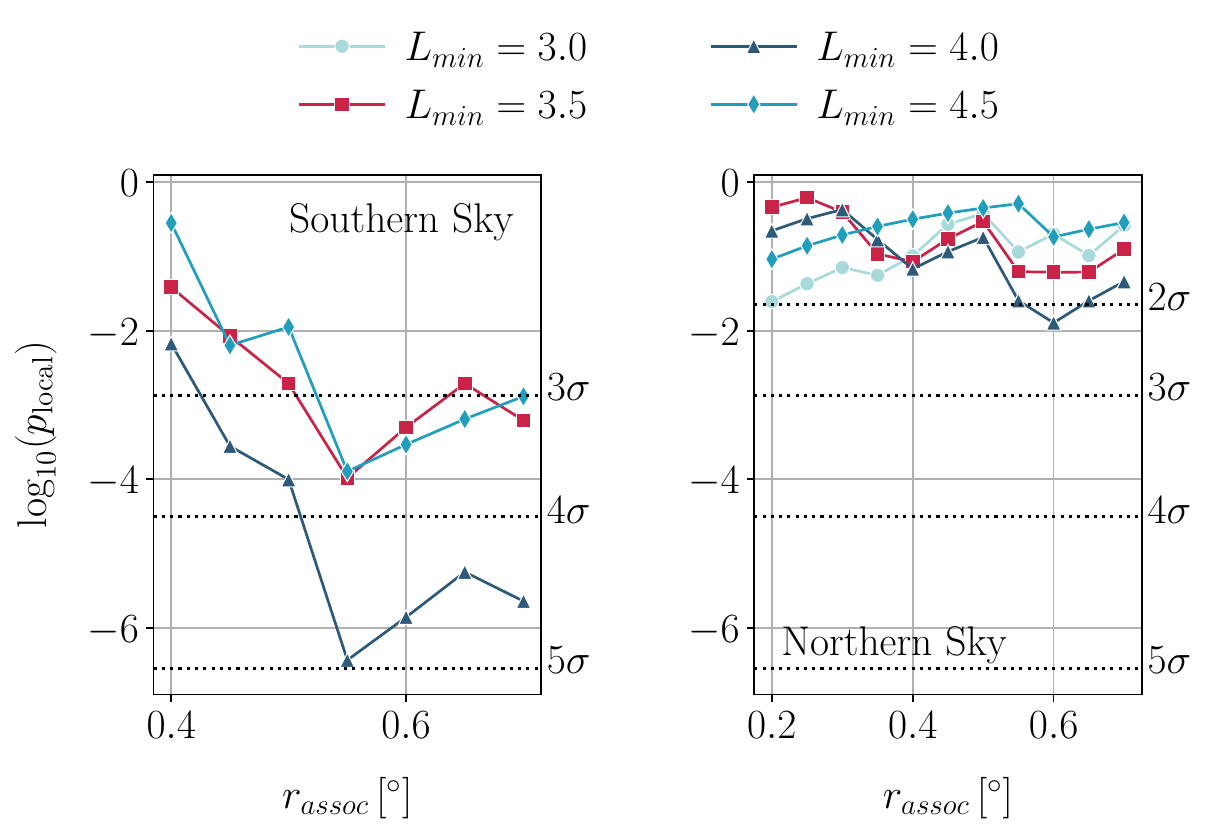}
    \caption{$p_{\rm local}$ for the 5BZCAT -- neutrino hotspot spatial correlation as a function of the association radius ($r_{\mathrm{assoc}}$) and for various minimum significance thresholds for the hotspots ($L_{\mathrm{min}}$) for the 7-year samples. The correlation analysis is performed both in the Southern (left) and in the Northern sky (right). The two panels also show the significance level corresponding to the number of Gaussian-equivalent standard deviations. The left-hand plot shows that we reproduce the results presented in 
    \cite{Buson2022a,Buson2022b}.}
    \label{fig:BZC_7yrs}
\end{figure}

\autoref{fig:BZC_7yrs} shows the 5BZCAT -- hotspot spatial correlation local (pre-trial) p-value vs.
$r_{\mathrm{assoc}}$ for the 7-year Southern sky (left panel) using the same $L_{\mathrm{min}}$ values as \cite{Buson2022a}.
This is almost exactly the same as their Fig. 1. We also derive the same minimum post-trial significance ($\sim 2 \times 10^{-6}$, $\sim 4.6 \sigma$) as \cite{Buson2022b} for $L_{\mathrm{min}} = 4.0$ and $r_{\mathrm{assoc}} = 0.55^{\circ}$.
The associated blazars are the same ten as those listed in Tab. 2 of \cite{Buson2022a}, i.e. five flat-spectrum radio quasars (FSRQs), three BL Lacs, and two blazars of uncertain type, with relatively large mean angular separations between the blazar and hotspot positions $\langle \psi \rangle \sim 0.40^{\circ}$ and mean redshift $\langle z \rangle \sim 1.3$.
All of them are by definition radio-detected and three are $\gamma$-ray sources. 

In the 7-year Northern sky, 112, 43, 18, and 7 neutrino hotspots have $L>L_{\rm{min}}$, with $L_{\rm{min}}$ ranging from 3.0 to 4.5 in steps of 0.5 (see \autoref{sec:corr_10yr}).
The correlation analysis results are compatible with the background (p-value $> 0.1\%$, that is equivalent to a significance lower than $3\sigma$), with a minimum local p-value $\sim 1\%$ (\autoref{fig:BZC_7yrs}, right panel) and a post-trial one of $\sim 10\%$.
Note that in this case, based on the results shown in \autoref{sec:method} and \autoref{sec:corr_10yr}, $r_{\mathrm{assoc}}$ starts from $0.2^{\circ}$ and the additional significance threshold $L_{\rm{min}}=3.0$ is used. 

\subsection{RFC -- 7-year hotspots correlation}
\begin{figure}[t!]
    \centering
    \includegraphics[width=\linewidth]{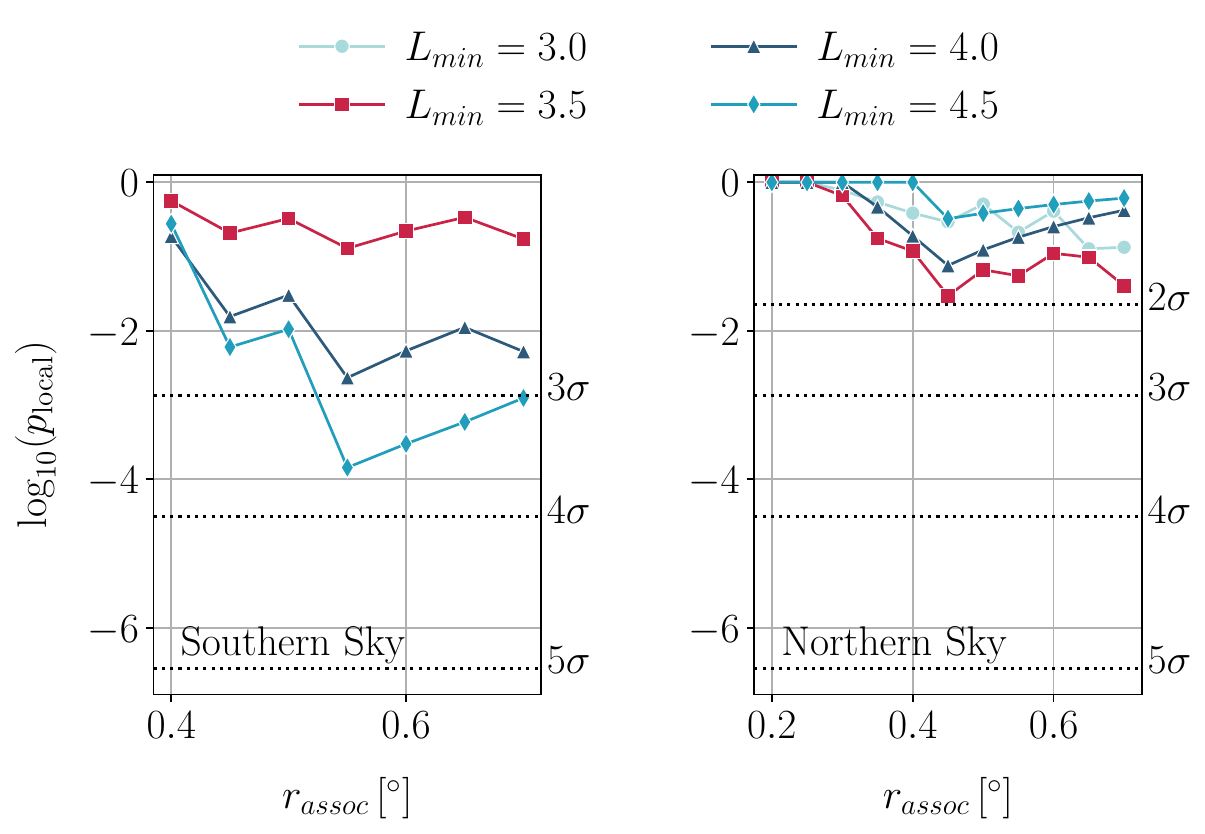}
    \caption{$p_{\rm local}$ for the RFC -- neutrino hotspot spatial correlation as a function of the association radius ($r_{\mathrm{assoc}}$) and for various minimum significance thresholds for the hotspots ($L_{\mathrm{min}}$) for the 7-year samples. The correlation analysis is performed both in the Southern (left) and in the Northern sky (right). The two panels also show the significance level corresponding to the number of standard deviations.}
    \label{fig:RFC_7yrs}
\end{figure}
\autoref{fig:RFC_7yrs} shows the RFC -- neutrino hotspot spatial correlation local (pre-trial) p-value vs.
$r_{\mathrm{assoc}}$ for the 7-year Southern sky (left panel; using the same $L_{\mathrm{min}}$ and $r_{\mathrm{assoc}}$ thresholds as in \citealt{Buson2022a}), which shows a minimum local p-value $1.4\times 10^{-4}$,
which becomes $6.9 \times 10^{-4}$ ($3.2\sigma$) post-trial, achieved for $L_{\mathrm{min}} = 4.5$ and $r_{\mathrm{assoc}} = 0.55^{\circ}$, the latter being the same value as for 5BZCAT.
This is easily explained by the fact that the RFC and 5BZCAT sources responsible for this minimum are the same.
Since for 5BZCAT $L_{\mathrm{min}}$ was 4.0, the associated blazars in this case are fewer, only five, i.e. four flat-spectrum radio quasars (FSRQs) and one blazar of uncertain type, with $\langle r_{\mathrm{assoc}}\rangle \sim 0.41^{\circ}$ and $\langle z \rangle \sim 1.8$. Only one of them is a $\gamma$-ray source.
However, similarly to the 5BZCAT case, in the 7-year Northern sky (\autoref{fig:RFC_7yrs}, right panel) the results are not significant, with a minimum local p-value $\sim 3\%$ and a post-trial one of $\sim 17\%$. 

\subsection{5BZCAT -- 10-year hotspots correlation}

\begin{figure}[t!]
    \centering
    \includegraphics[width=\linewidth]{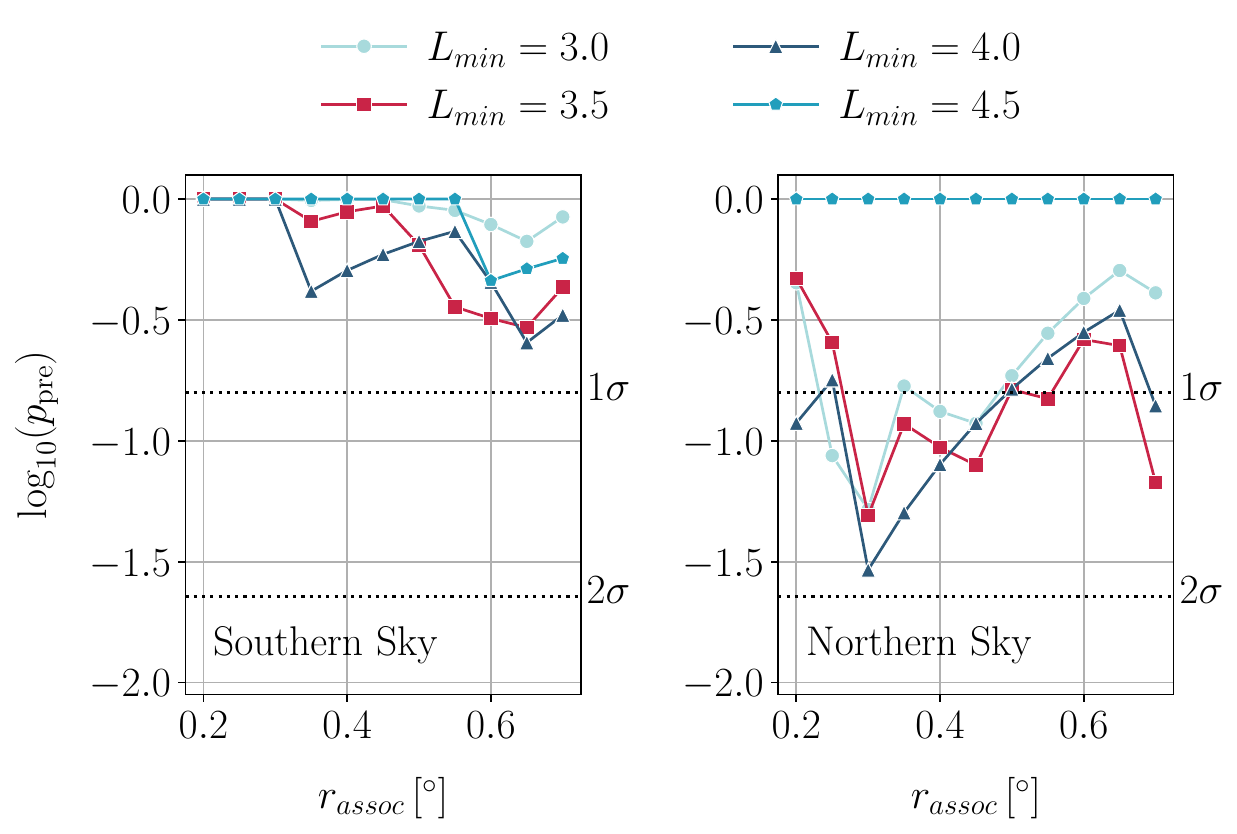}
    \caption{$p_{\rm local}$ for the 5BZCAT -- neutrino hotspot spatial correlation as a function of the association radius ($r_{\mathrm{assoc}}$) and for various minimum significance thresholds for the hotspots ($L_{\mathrm{min}}$) for the 10-year neutrino sample. The correlation analysis is performed both in the Southern (left) and in the Northern sky (right). The two panels also show the significance level corresponding to the number of standard deviations.}
    \label{fig:BZC_10yrs}
\end{figure}

In the 10-year Southern sky, 149, 58, 18, and 7 hotspots have $L>L_{\rm{min}}$ with $L_{\rm{min}}$ in [3.0, 3.5, 4.0, 4.5],  while the Northern sky has 95, 36, 7, and 2 hotspots with significance above the same set of thresholds. We note that while \cite{10yearPS} reported a local significance $L=3.7$ at the location of the blazar TXS~0506+065, this object has a lower significance $L=2.7$ in this analysis. This is interpreted as due to the detector response matrix, which averages the detector smearing of the incoming neutrinos in a bin ranging from $-10^{\circ}$ to $+10^{\circ}$, where IceCube's data selection and processing change (see \autoref{app:skymap} for more details). Having a significance below the minimum tested threshold of $L_{\rm{min}}=3.0$, TXS~0506+056 is not included in our correlation analysis. However, even the inclusion of a hotspot coincident with TXS~0506+056 does not make the correlation of either of the two tested catalogs incompatible with a chance coincidence (see \autoref{app:TXS}).

\autoref{fig:BZC_10yrs} shows the 5BZCAT -- hotspot spatial correlation local (pre-trial) p-value vs.
$r_{\mathrm{assoc}}$ for the 10-year Southern (left panel) and Northern (right panel) skies.
The results are not significant, with minimum local (post-trial) p-values $\sim 25\%$ ($\sim 75\%$) and $\sim 3\%$ ($\sim 19\%$), respectively. 

\subsection{RFC -- 10-year hotspots correlation}

\begin{figure}[t!]
    \centering
    \includegraphics[width=\linewidth]{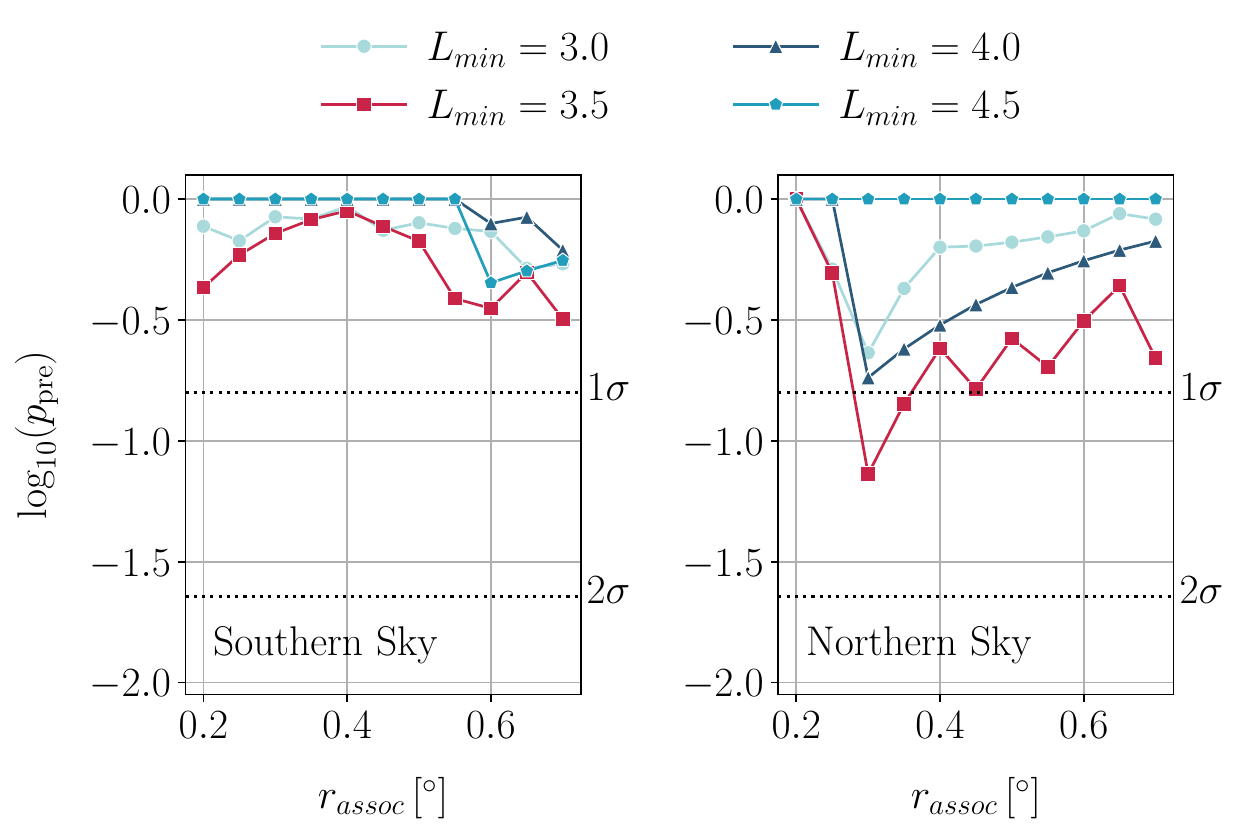}
    \caption{$p_{\rm local}$ for the RFC -- neutrino hotspot spatial correlation as a function of the association radius ($r_{\mathrm{assoc}}$) and for various minimum significance thresholds for the hotspots ($L_{\mathrm{min}}$) for the 10-year neutrino sample. The correlation analysis is performed both in the Southern (left) and in the Northern sky (right). The two panels also show the significance level corresponding to the number of standard deviations.}
    \label{fig:RFC_10yrs}
\end{figure}

\autoref{fig:RFC_10yrs} shows the RFC -- hotspot spatial correlation local (pre-trial) p-value vs.
$r_{\mathrm{assoc}}$ for the 10-year Southern (left panel) and Northern (right panel) skies.
Again the results are not significant, with minimum local (post-trial) p-values $\sim 32\%$ ($\sim 80\%$) and $\sim 7\%$ ($\sim 34\%$) respectively. 

The results from the different correlation analyses described in this Section are summarized in \autoref{tab:results}.

\begin{table}[]
    \centering
    \caption{Post-trial p-values for the different positional correlation analyses performed in this work.}
    \begin{tabular}{ l|c|c|c|c| }
          &  \multicolumn{2}{c|}{Southern sky} & \multicolumn{2}{c|}{Northern sky}\\
        \cline{2-5}
                & 7-yr $\upnu$  &  10-yr $\upnu$  &  7-yr $\upnu$  &  10-yr $\upnu$  \\
        \hline
        5BZCAT  &  2.5$\times10^{-6}$  &  0.75  &  0.10  &  0.19  \\
        \hline
        RFC     &  6.9$\times10^{-4}$  &  0.80  &  0.17  &  0.34  \\
        \hline
    \end{tabular}
    \label{tab:results}
\end{table}

\section{Discussion}\label{sec:discussion}
\subsection{5BZCAT}\label{sec:discussion_BZCAT}
We have reproduced the very significant result of \cite{Buson2022a,Buson2022b} on the neutrino hotspot -- 5BZCAT blazar correlation using 7 years of IceCube data in the Southern sky.
However, this apparently strong result is not robust since when applying the same type of analysis to the 7-year Northern sky the correlation becomes insignificant.
Moreover, when using the larger and improved 10-year IceCube sample covering the whole sky, no correlation is found in either hemisphere. 

\begin{figure}[h]
    \centering
    \includegraphics[width=.45\textwidth]{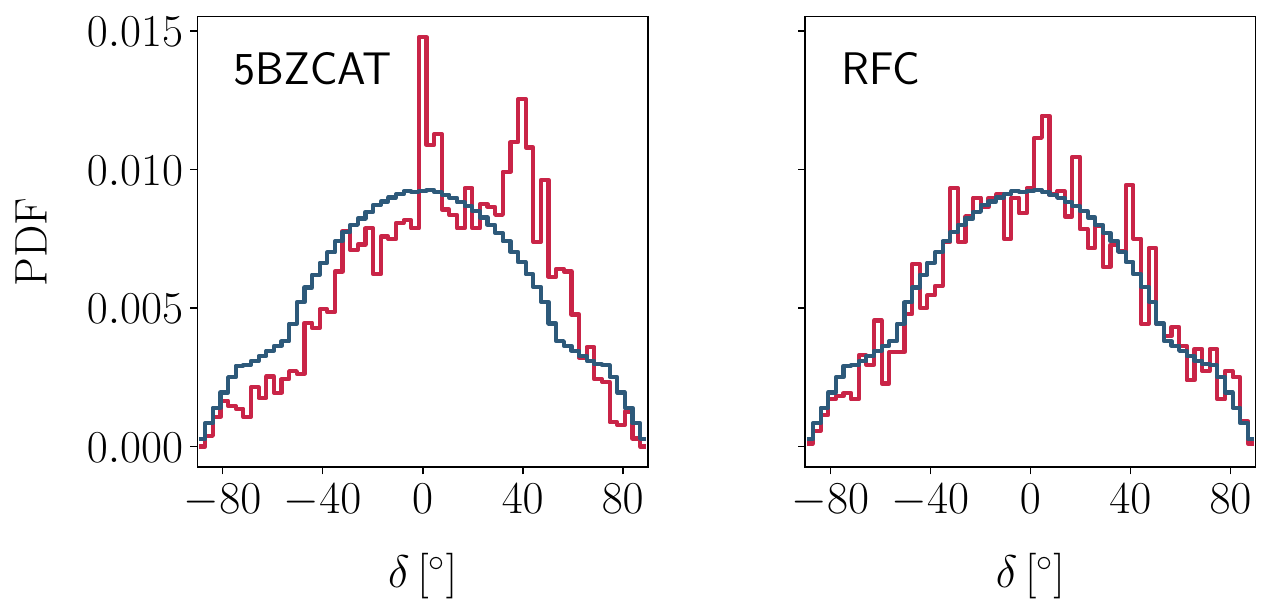}
    \caption{Declination distribution of the sources in the 5BZCAT (left) and in the RFC (right) in red compared to the cosine distribution in blue. Note that the expected uniform distribution is not a perfect cosine because the Galactic plane is removed, as done in the source catalogs.}
    \label{fig:declinations}
\end{figure}

Given the nature of the 5BZCAT catalog (\autoref{blazar_sample}), its utilization for a statistical test in place of other available flux-limited samples, such as the RFC catalog, is an arguable choice.
Moreover, being nine years old, 5BZCAT is also by now quite incomplete: for example, while the latest incremental {\it Fermi}-4LAT catalog \citep{4LAC-DR3} contains more than 3,700 blazars, 5BZCAT includes only $\sim 1,000$ $\gamma$-ray emitting sources.
Indeed, the left panel of \autoref{fig:declinations} shows that the declination distribution of 5BZCAT is very different from uniform, as the Southern hemisphere is severely under-sampled while there are two notable excesses at $\delta \sim 0^{\circ}$ and $\delta \sim 40^{\circ}$.
The former point is well-known and due to a radio-optical identification bias towards the Northern sky \citep[e.g.][]{BZCAT5}.
We test the hypothesis that the declination distribution of the sources is a realization of the uniform distribution (excluding the Galactic plane) using the Kolmogorov-Smirnov (KS) and Wilcoxon-Mann-Whitney (WMW) tests. Both tests result in strong incompatibility, with a significance larger than 8$\sigma$.

This situation is similar to that of the \cite{Auger2008}, in which the authors had found an excess (chance probability $1.7 \times 10^{-3}$) in the correlation of ultra-high-energy cosmic rays (UHECRs) above 57 EeV lying within $3.1^{\circ}$ of sources closer than 75 Mpc in the AGN catalog of \cite{Veron2006} 
(this catalog is also a very inhomogeneous list of sources).
However, the fraction of arrival directions that were fulfilling the criteria above went from $69^{+11}_{-13}$\% (to be compared with 21\% expected for isotropic UHECRs), down to $38^{+7}_{-6}$\% \citep{Abreu2010}, and then $33\pm5$\% and finally $28.1^{+3.8}_{-3.6}$\% \citep{Aab2015}.
To quote the latter paper: ``the high level of correlation found initially was probably affected by a statistical fluctuation''.
We have shown here that this is also the case for the IceCube--5BZCAT correlation.

The distribution of the angular distance between the hotspots and the ten associated sources in \cite{Buson2022a} deserves some scrutiny.
The mean offset is $0.4^{\circ}$, which is $\sim 3$ times larger than the median localization capability of the point source analysis used here to analyze the 10-year neutrino sample in the Southern sky (see \autoref{sec:corr_10yr} and \autoref{fig:localization}). 7/10 hotspots are found at an angular distance $\psi \geq 0.4^{\circ}$ from the source they are associated with.
The binomial probability that this happens using our point source analysis on the 10-year sample is $\sim 2\times 10^{-6}$. With respect to the 7-year data used by \cite{Buson2022a,Buson2022b}, the average angular resolution on the single event in the 10-year dataset has improved by $\sim 10\%$ for the last six years only, while the first four years are treated exactly as before.
Hence, the localization capability would only slightly worsen when applied to the 7-year sample (for which no neutrino events are publicly available).
However even considering an extremely bad localization performance, e.g. assuming the angular resolution of the source localization with the 7-year neutrino sample is a factor of 2 larger than with the 10-year sample,
the binomial probability defined above would still be extremely small, $\sim 10^{-3}$. This makes the observed correlation inconsistent with a hypothetical signal and provides additional and independent evidence that the correlation claimed by \cite{Buson2022a,Buson2022b} could be due to the blazar sample used and statistical fluctuations.

\subsection{RFC}\label{sec:discussion_RFC}
\cite{Plavin2021} analyzed the 7-year IceCube neutrino $L$ map against the positions of the (blazar-dominated) RFC sources with $S^{\rm VLBI}_{\rm 8~GHz} \ge 150$ mJy and $\delta > -5^{\circ}$.
Since they noticed a trend of larger $L$ values at higher $\delta$, they first derived medians and then used normalized values $L_{\rm norm} = L - L_{\rm med}$, where $L_{\rm med}$ corresponds to the pixel’s $\delta$ and is subtracted from the original value in each pixel.
They then selected their blazar-dominated sample with VLBI flux density higher than a given $S_{\rm min}$, extracted $L_{\rm norm}$ values at their positions, and took the median of these values as the test statistic.
They finally tested whether this value was higher than expected by chance by performing the same calculations on the RFC positions randomly shifted in right ascension 100,000 times. 
The minimum local p-value of $4 \times 10^{-4}$ was obtained for $S_{\rm min} = 0.33$ Jy, while the post-trial one turned out to be $3 \times 10^{-3}$, i.e. a Gaussian-equivalent of $3.0\sigma$. 

\autoref{fig:declinations} plots the RFC $\delta$ distribution (right panel).
WMW and KS tests on the compatibility of the declination distribution of the sources with the reference uniform distribution that excludes the Galactic plane return p-values $\sim 2 - 4\%$ respectively, proving that the sources in the RFC catalog are distributed much more uniformly in the sky than the ones in the 5BZCAT, although still with a small Northern sky bias.
However, the p-values for the Northern sky alone, which was the one used by \cite{Plavin2021}, are $\sim 39 - 61\%$, i.e. perfectly consistent with a uniform distribution. 

Using the analysis technique described in \autoref{sec:method}, we have derived a moderately significant ($6.9 \times 10^{-4}$ post-trial, equivalent to $\sim 3.2\sigma$) result for the association between the 7-year Southern sky and the RFC but, contrary to \cite{Plavin2021}, we obtained an insignificant result in the 7-year Northern sky. 
The correlation in the Southern sky comes from the association of 5 sources for which the median angular distance from their respective hotspots is $0.41^{\circ}$.
Four of these objects have angular offsets $\psi \geq 0.41^{\circ}$.
We can repeat the exercise already done for the blazars correlated in the 5BZCAT and calculate the binomial probability that this happens, which is $5\times 10^{-4}$ if we use the distribution in \autoref{fig:localization} (right panel) and $7\times 10^{-3}$ in the degraded scenario where the resolution of the source localization is doubled.

Moreover, as for 5BZCAT, no correlation was found in either hemisphere when using the larger and improved 10-year IceCube sample covering the whole sky.
Hence, we come to the conclusion that our RFC result in the Southern sky is also most likely a statistical fluctuation.

We also note that \cite{Abbasi2023} investigated 
the results of \cite{Plavin_2020}, who found a 
$3.1\sigma$ correlation between a (by now outdated) list of 56 IceCube high-energy neutrinos and a previous version of the RFC catalog. 
They found that adding more information on the neutrino events and more data overall by using IceCat-1 \citep{IceCat1}, the very recent catalog of likely astrophysical neutrino track-like events, made the \cite{Plavin_2020} results compatible with a statistical
fluctuation.

\subsection{The blazar contribution to IceCube neutrinos}\label{sec:discussion_blazar}
Where does all of the above leave us in terms of blazars as IceCube neutrino sources? It has been shown by several papers that their contribution cannot be predominant, especially at lower energies ($\lesssim 0.5$ PeV: e.g. \citealt{Padovani_2015,Aartsen_2016,Oikonomou_2022} and references therein) and might be restricted to specific blazar sub-classes \citep[e.g.][]{Padovani2016,Giommmi2021,Paiano_2023} or sources in a flaring state \citep[e.g.][]{Murase2018ApJ}.
The IceCube collaboration has also derived spectral model-dependent upper limits on the TeV--PeV neutrino emission from blazars, finding that:
1. {\it Fermi}-2LAC blazars \citep[detected in the $0.1 - 100$ GeV range:][]{2LAC} cannot contribute more than $\sim7\%$ to the diffuse astrophysical neutrino flux from \cite{IceCube2015ApJ} assuming an $E^{-2.5}$ neutrino energy spectrum \citep{IceCube2LAC2017ApJ};
2. blazars from the 3FHL catalog \citep[detected in the $10 - 2000$ GeV range:][]{3FHL} cannot contribute more than $\sim10\%$ to the diffuse flux from \cite{Haack2017ICRC} 
assuming an $E^{-2.19}$ neutrino energy spectrum \citep{2019ICRCHuber};
3. blazars from the 1FLE catalog \citep[detected in the very low-energy range $30 - 100$ MeV:][]{1FLE}  cannot contribute more than $\sim1\%$ to the diffuse flux from \cite{IceCubeDiffuse2022ApJ} assuming an $E^{-2.37}$ neutrino energy spectrum \citep{IceCube2022ApJ}.
These constraints depend strongly on the assumed power-law spectrum.
However, the neutrino emission expected from blazars is very peaked and not power-law-like \citep[e.g.][]{Oikonomou_2022}. 

\cite{Padovani2022}, based on the $3.3\sigma$ excess in the Northern sky due to significant p-values in the directions of NGC 1068 and three blazars \citep{10yearPS}, namely TXS~0506+056, PKS~1424+240, and GB6~J1542+6129, with very high radio and $\gamma$-ray powers, have suggested that blazars of the type IceCube has associated with neutrinos constitute at most $1\%$ of the $\gamma$-ray selected population.
\cite{Abbasi2023} came exactly to the same conclusion using a completely different approach, namely by estimating the fraction of {\it Fermi} 4LAC-DR2 blazars compatible with the signal they detected.
If indeed only a small fraction of the $\gamma$-ray and radio-selected blazars are sources of high-energy neutrinos, then any cross-correlation between IceCube events and large blazar samples of the type studied in this paper is bound to wash out a possible signal. 

\section{Conclusions}
We have reproduced and extended the work of \cite{Buson2022a,Buson2022b}, who tested the association of blazars in the 5BZCAT catalog with an IceCube neutrino map covering 7 years of observations in the Southern sky, to the Northern sky and to a $\sim 60\%$ larger and improved, recently released all-sky IceCube dataset covering 10 years.
While we managed to reproduce the $\sim 4.6\sigma$ equivalent p-value of \cite{Buson2022a,Buson2022b}, only insignificant results were obtained in the 7-year Northern sky and in the 10-year data for both Northern and Southern hemispheres.
The original finding appears then to be only a statistical fluctuation, which might be partly due to the nature of 5BZCAT, which is not a flux-limited, well-defined catalog but just a (by now 9-year-old) list of sources detected in various bands.

We have also applied the same methodology to a complete, blazar-dominated AGN sample with 8 GHz VLBI flux density $\ge 150$ mJy.
In this case, we have derived a $\sim 3.3\sigma$ equivalent result in the 7-year Southern sky but, contrary to \cite{Plavin2021} who used a different approach but the same catalog, got an insignificant result in the 7-year Northern sky.
However, as for 5BZCAT, no correlation was found in the 10-year whole sky. 

All of the above does not exclude a blazar contribution to the IceCube signal, which is supported by the neutrino association with TXS\,0506+056, but is consistent with a scenario where only a small fraction of the $\gamma$-ray- and radio-selected blazars are high-energy neutrino sources, in agreement with previous works.

\begin{acknowledgments}
We thank the referee for the insightful and very detailed comments. C.B. thanks S. Coenders and M. Wolf for their help in developing parts of the analysis used in this work, and M. Karl and H. Niederhausen for proofreading this manuscript.
This work is supported by the Deutsche Forschungsgemeinschaft through grant SFB 1258 ``Neutrinos and Dark Matter in Astro- and Particle Physics''.
\end{acknowledgments}

%

\vspace{5mm}
\facilities{The IceCube Neutrino Observatory.}


\software{\textsc{SkyLLH}\footnote{https://github.com/icecube/skyllh/tree/master/skyllh}\citep{SkyLLHICRC2023}, \textsc{healpy} \citep{Zonca2019}}



\onecolumngrid
\appendix

\section{The likelihood analysis}\label{app:llh}
We search the entire sky for continuous neutrino emission using 10 years of muon tracks recorded and processed by the IceCube collaboration.
We use an adaptation of the unbinned maximum-likelihood method described by \cite{Braun2008} to search for neutrino point-like sources.
We search for clusters of events in the sky while also using the information on their energy to further discriminate from random accumulations produced by the background. Hence, the statistical model is a mixture of a signal and a background model so that the data can be described by one of the two following hypotheses:
\begin{itemize}
    \item $H_0$: the recorded events only consist of background muons, produced in the atmosphere or by the astrophysical diffuse neutrino flux;
    \item $H_1$: the recorded events are a mixture of background muons and of an additional component originating from an astrophysical point-like neutrino source, which emits with some strength $n_{\rm{s}}$ according to a power-law energy spectrum $\propto E^{-\gamma}$.
\end{itemize}

The test-statistic $(TS)$ is the negative logarithm of the likelihood-ratio:
\begin{equation}\label{eq:TS}
    TS=-2\log{\frac{\mathcal{L}(H_0|Data)}{\mathcal{L}(H_1|Data)}},
\end{equation}
which is maximized with respect to the 2 free parameters in the signal model, $n_{\rm{s}}\geq0$ and $\gamma \in [1,5]$.

\begin{figure}
    \centering
    \includegraphics[width=.95\linewidth]{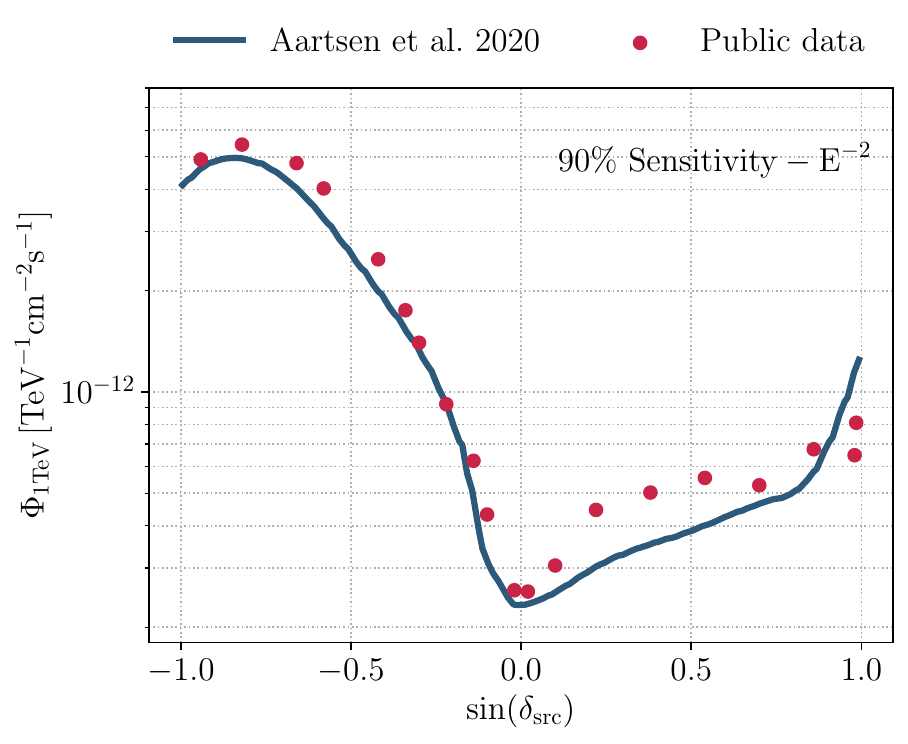}
    \caption{Comparison of the sensitivity to an E$^{-2}$ astrophysical neutrino flux across the sky for the analysis used in this work and the one published in \cite{10yearPS}.}
    \label{fig:sensitivity}
\end{figure}

The null hypothesis $H_0$ is a particular case of the alternative hypothesis $H_1$, i.e., $H_1$ reduces to $H_0$ when the point-like emission strength is zero ($n_{\rm{s}}$=0). Therefore the two hypotheses are nested, and the likelihood of observing the data given the model is

\begin{equation}
    \mathcal{L}(n_{\rm{s}}, \gamma|\vec{x})= \prod_{i}^{N}\left[\frac{n_{\rm{s}}}{\mathrm{N}}\mathcal{S}(x_i|\gamma) + \left(1-\frac{n_{\rm{s}}}{\mathrm{N}}\right)\mathcal{B}(x_i)\right],
\end{equation}
where the likelihood is unbinned, so the product is over all $N$ events $\{x_i\}$ in the sample, and $\mathcal{S}$ and $\mathcal{B}$ are the signal and background probability density functions (pdf), respectively. Both depend on the reconstructed observables for the $i$-th event, $x_i=(E_{\upmu,i}, \vec{d}_{\upmu,i}, \sigma_i)$, where $E_{\upmu}$ is the muon energy, $\vec{d}_{\upmu}$ is the muon direction and $\sigma$ is the uncertainty on the reconstructed direction\footnote{By applying an energy-dependent correction factor to the reconstructed angular uncertainty, the $\sigma$ variable reported in the 10-year data release already takes into account the kinematic opening angle between the parent neutrino and the muon \citep{ICdatarelease}, which becomes important at energies $\lesssim 1$~TeV \citep[e.g.][and references therein]{2021PhDGlauch}.}. The signal pdf additionally features a conditional dependence on the spectral index, $\gamma$, of the energy distribution. Each of the two pdfs can be separated into a spatial pdf and an energy pdf.
The spatial and energy distributions of background events as well as the spatial distribution of signal events (a 2D Gaussian centered at the assumed source location and with standard deviation equal to the $\sigma$ of the event) are the same used in IceCube's typical point-source analysis \citep[e.g.][]{2019PhDCarver}. The energy pdf in our signal likelihood is the only term that differs from the typical IceCube analysis. In fact, while IceCube analyses usually define a signal energy pdf conditional on the reconstructed muon declination, our signal energy pdf conditionally depends on the assumed source declination\footnote{Dependencies on the right ascension can be ignored due to the position of the detector at the South Pole \citep[e.g.][]{2016PhDCoenders}.}.
This modification is needed to cope with the absence of the reconstructed muon declination in the detector response matrix, which is used to construct the signal energy pdf as
\begin{equation}\label{eq:energy_pdf}
    P(E_{\upmu}|\delta_{\upnu},\gamma) = \int dE_{\upnu}\,P(E_{\upmu}|E_{\upnu},\delta_{\upnu})\,P(E_{\upnu}|\gamma)\,P(E_{\upnu}|\delta_{\upnu}).
\end{equation}
The terms of the energy pdf in \autoref{eq:energy_pdf} are:
\begin{itemize}
    \item $P(E_{\upmu}|E_{\upnu},\delta_{\upnu})$ is the probability of the muon energy $E_{\upmu}$ given the parent neutrino energy $E_{\upnu}$ and declination $\delta_{\upnu}$, and it is obtained from the detector response matrix;
    \item $P(E_{\upnu}|\gamma)$ is the probability of the neutrino energy given a power-law energy flux with spectral index $\gamma$;
    \item $P(E_{\upnu}|\delta_{\upnu})$ is the probability of detecting a neutrino of energy $E_{\upnu}$ from the direction $\delta_{\upnu}$, calculated from the tabulated effective areas.
\end{itemize}
To compare our likelihood analysis to the one in \cite{10yearPS}, we compare the respective sensitivities to an astrophysical $E^{-2}$ neutrino flux simulated at various declinations across the sky\footnote{For a definition of the sensitivity flux see e.g. \cite{2022PhDKarl}.}. The events for the signal injection are sampled from the detector response matrix. The analysis used in this work is sensitive to an $E^{-2}$ astrophysical flux 20--50\% higher than the one that the analysis in \cite{10yearPS} would detect, depending on the declination, as shown in \autoref{fig:sensitivity}.

\section{The 10-year neutrino p-value map}\label{app:skymap}
\begin{figure}
    \centering
    \includegraphics[width=.95\linewidth]{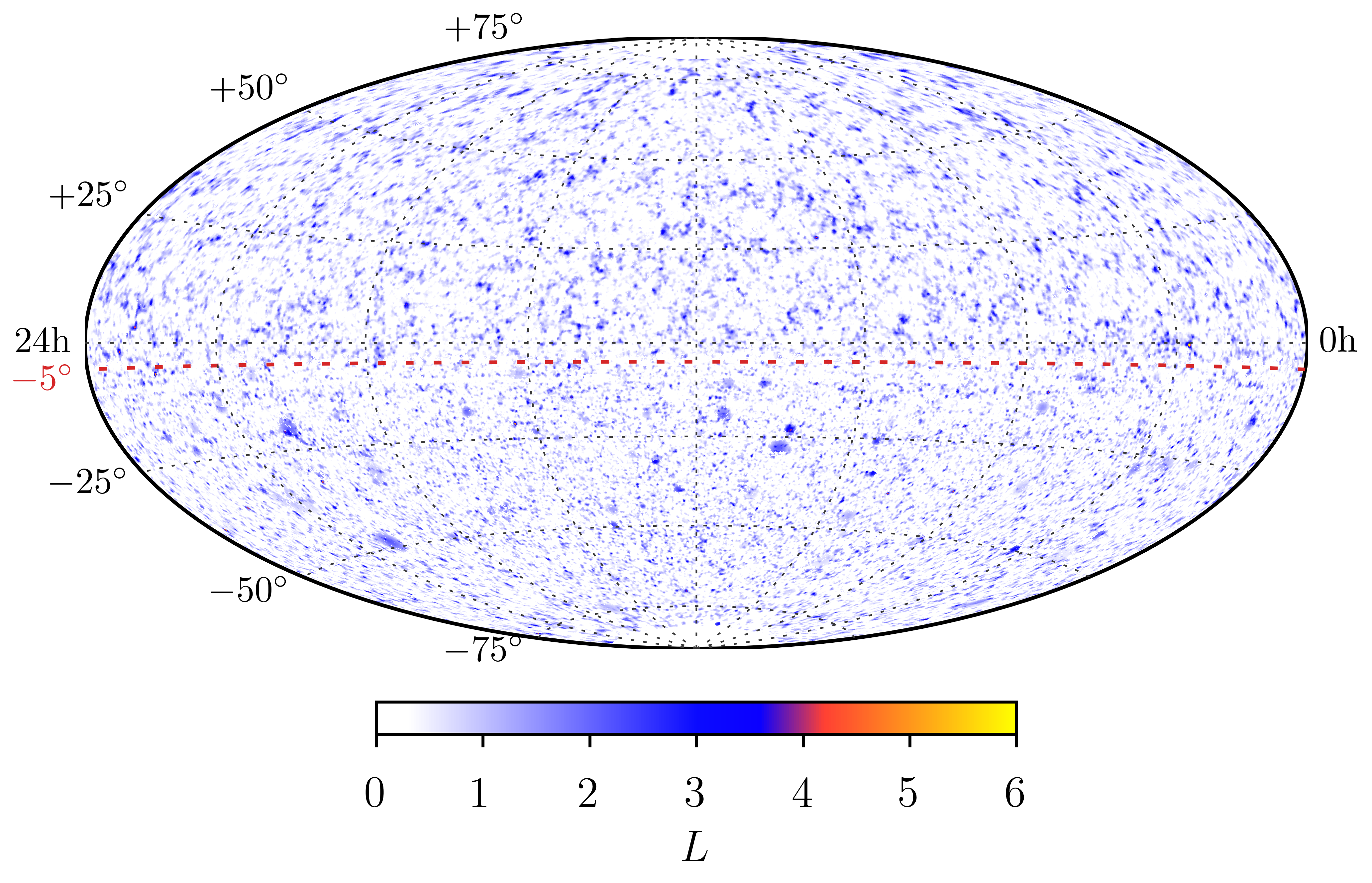}
    \caption{$L$ sky map of the point source likelihood search in the Northern and Southern hemispheres. The map is shown in equatorial coordinates (right ascension on the horizontal axis and declination on the vertical one) on a Hammer-Aitoff projection. The color scale indicates the $L$ values obtained from the maximum likelihood-ratio analysis performed at each pixel in the sky. The dashed red line indicates the horizon, separating the Northern and Southern skies at $-5^{\circ}$.}
    \label{fig:skymap}
\end{figure}

To calculate the 10-year neutrino $L$ [i.e., $-\log_{10}$(p-value)] map, we divide the sky in a grid of equivalent pixels of $\sim (0.11^{\circ})^2$ constructed using a HEALPix \citep{Gorski2005} projection with \textsc{Nside} parameter 512. At each of them, we maximize the TS (\autoref{eq:TS}) and calculate the $L$ value as the negative logarithm of the probability of obtaining a TS larger than the observed one from background pseudo-experiments, generated by scrambling the experimental data in right ascension multiple times \citep[e.g.][]{2019PhDCarver}. Thus the $L$ value at each pixel represents the probability that the data belong to the background-only hypothesis. The neutrino $L$ sky map computed in this work is shown in \autoref{fig:skymap}.

We check the local significance at the locations of the four most significant sources contributing to the 3.3$\sigma$ excess from a binomial test on a list of candidate neutrino sources published in \cite{10yearPS}. We find $L=4.41$ for NGC~1068, $L=2.70$ for TXS~0506+056, $L=3.15$ for PKS~1424+240 and $L=2.52$ for GB6~J1542+6129. The respective $L$ values published by the IceCube collaboration are $L=4.74$, $L=3.72$, $L=2.80$, and $L=2.74$. Differences are within $\sim0.2$ Gaussian-equivalent standard deviations, except for the blazar TXS~0506+056 which has a local p-value an order of magnitude larger than the one reported by the IceCube collaboration using the same dataset (equivalent to a difference of $\sim0.66\sigma$).
In the Southern sky, we find a local significance of $L=4.83$ for the published hottest spot, located at right ascension $350.2^{\circ}$ and declination $-56.5^{\circ}$. This is slightly lower than the value reported by IceCube of $L=5.3$ but still compatible within $\sim0.2$ standard deviations.
The differences in the observed p-values, as well as in the sensitivity fluxes (\autoref{fig:sensitivity}), can have various causes, from the difference in the energy term of the spatial likelihood to the coarseness of the information provided in the public detector response matrices. Indeed, it is worth recalling here that these matrices only have 3 bins in neutrino declination, covering the sky from $-90^{\circ}$ to $-10^{\circ}$, from $-10^{\circ}$ to $10^{\circ}$, and from $10^{\circ}$ to $90^{\circ}$. Especially at the horizon (where TXS~0506+056 is located), the detector response is averaged over a declination range where the detector behavior changes significantly: the neutrino declination bin includes the border between Southern and Northern sky, at $-5^{\circ}$,  where data processing and selection cuts change \citep[e.g.][and references therein]{10yearPS}.

\section{Including the blazar TXS~0506+056 in the correlation analysis}\label{app:TXS}
\begin{figure}[h]
    \centering
    \includegraphics[width=\linewidth]{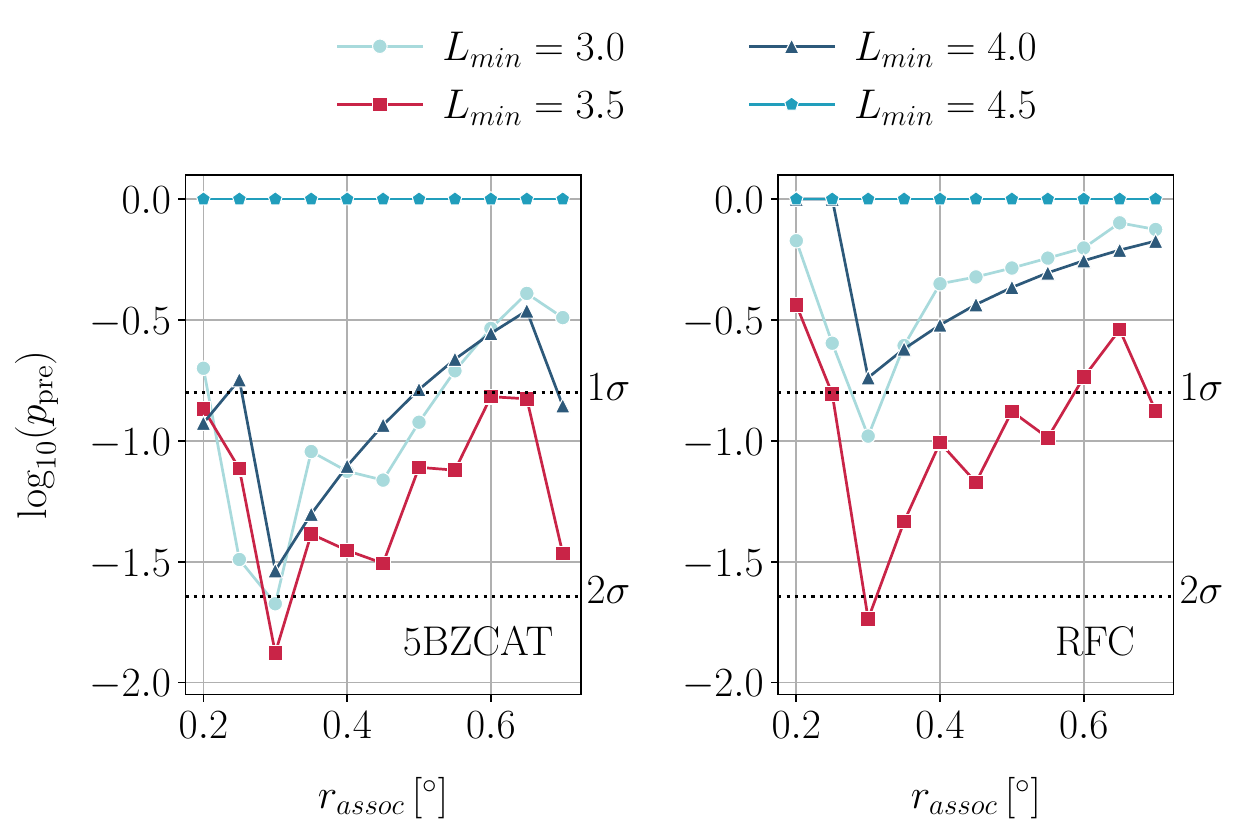}
    \caption{$p_{\rm local}$ for the blazar -- neutrino hotspot spatial correlation as a function of the association radius ($r_{\mathrm{assoc}}$) and for various minimum significance thresholds for the hotspots ($L_{\mathrm{min}}$) for the 10-year Northern neutrino sky. A hotspot with $L_{\rm{min}}=3.7$ at the location of TXS~0506+056 was added to the hotspot list. The correlation analysis uses the 5BZCAT (left) and the RFC (right) catalogs. The two panels also show the significance level corresponding to the number of Gaussian-equivalent standard deviations.}
    \label{fig:north_wTXS}
\end{figure}

The significance of the well-known blazar TXS~0506+056 in the 10-year neutrino p-value map produced in this work is smaller than the one reported by \cite{10yearPS}. Having $L=2.7$, it does not pass the threshold $L_{\rm{min}}=3.0$ for being selected as an interesting hotspot for the correlation analysis. As a result, even though this source is included in both the 5BZCAT and RFC catalogs, the hotspot associated with it is not correlated with the two catalogs. As an a-posteriori check, we test the impact of the absence of this hotspot on our analysis. 
We add to the list of our neutrino hotspots in the Northern hemisphere an additional hotspot at the pixel corresponding to the location of the blazar and with the local significance reported in \cite{10yearPS}, i.e. $L=3.7$. We then run again the cross-correlation analysis in the Northern sky with both the 5BZCAT and the RFC catalogs. The results of this cross-check are shown in \autoref{fig:north_wTXS}. The best-fit $L_{\rm{min}}$ changes from 4.0 to 3.5 when using the 5BZCAT, while it stays at 3.5 for the RFC catalog. The pre-trial p-value of the correlation slightly decreases in both cases, going from 2.9\% to 1.3\% for the 5BZCAT and from 7.9\% to 1.8\%, but remaining fully compatible with the background expectation.


\bibliography{main}{}
\bibliographystyle{aasjournal}



\end{document}